\documentstyle[twoside,fleqn,espcrc2]{article}

\newcommand{\nc}{\newcommand}
\nc{\be}{\begin{equation}}
\nc{\ee}{\end{equation}}
\nc{\bea}{\begin{eqnarray}}
\nc{\eea}{\end{eqnarray}}

\nc{\eqn}[1]{{(\ref{#1})}}
\nc{\cA}{{\cal A}}
\nc{\cB}{{\cal B}}
\nc{\cC}{{\cal C}}
\nc{\cD}{{\cal D}}
\nc{\cE}{{\cal E}}
\nc{\cF}{{\cal F}}
\nc{\cG}{{\cal G}}
\nc{\cH}{{\cal H}}
\nc{\cI}{{\cal I}}
\nc{\cJ}{{\cal J}}
\nc{\cK}{{\cal K}}
\nc{\cL}{{\cal L}}
\nc{\cM}{{\cal M}}
\nc{\cN}{{\cal N}}
\nc{\cO}{{\cal O}}
\nc{\cP}{{\cal P}}
\nc{\cQ}{{\cal Q}}
\nc{\cR}{{\cal R}}
\nc{\cS}{{\cal S}}
\nc{\cT}{{\cal T}}
\nc{\cU}{{\cal U}}
\nc{\cV}{{\cal V}}
\nc{\cW}{{\cal W}}
\nc{\cX}{{\cal X}}
\nc{\cY}{{\cal Y}}
\nc{\cZ}{{\cal Z}}

\newcommand{\AmS}{{\protect\the\textfont2
  A\kern-.1667em\lower.5ex\hbox{M}\kern-.125emS}}

\title{
  \begin{flushright} {\small HD-THEP-96-32} \end{flushright}
\vspace{-0.5cm}
Linked Cluster Expansions on non-trivial
topologies}

\author{T. Reisz\thanks{Heisenberg fellow, e-mail address
        t.reisz@thphys.uni-heidelberg.de}\address{Institut
        f\"ur Theoretische Physik,\\ 
	Universit\"at Heidelberg, \\ 
        Philosophenweg 16, \\
	D-69120 Heidelberg, Germany}
}

\begin{document}

\begin{abstract}
Linked cluster expansions provide a useful tool both for analytical and
numerical investigations of lattice field theories. 
The expansion parameter is the interaction strength
fields at neighboured lattice sites are coupled.
They result into convergent series
for free energies, correlation functions and susceptibilities.
The expansions have been generalized to field 
theories at finite temperature
and to a finite volume.
Detailed information on critical behaviour can be extracted from
the high order behaviour of the susceptibility series.
We outline some of the steps by which the 20th order is achieved.
\end{abstract}

\maketitle

\section{SERIES EXPANSIONS}

Realistic physical models evolve enor\-mous com\-plexity and 
in many instances force
us to use approximations or expansions about solvable cases.
Convergent expansions allow for
precise quantitative statements
within their range of convergence.
Under general conditions
they are suitable for an investigation of
phase transitions and of critical phenomena.
The price to be paid is that normally high orders have to be
computed in order to reveal collective behaviour, 
and special techniques have to be developed and applied.

The linked cluster expansion (LCE) amounts to a 
convergent hopping parameter
expansion applied to the free energy or connected correlation
functions of lattice models \cite{wortis}.
To be specific, let us consider the $O(N)$ symmetric scalar model
defined on the hypercubic lattice $\Lambda$ by the action
\be \label{action}
   S(\Phi,v) =  \sum_{x\in\Lambda} 
   {V}(\Phi_x) - \frac{1}{2}\;
   \sum_{x,y\in\Lambda}\sum_{a=1}^N \Phi^a_x v_{xy}\Phi^a_y
\ee
with
\be \label{hopping}
   v_{xy} \; = \; \left\{ 
   \begin{array}{r@{\qquad ,\quad} l }
    2\kappa\; & {\rm x,y \; nearest\; neighbour,} \\
    0 & {\rm otherwise,}
   \end{array} \right.
\ee
and lattice site action ${V}(\Phi_x)$ appropriately
boun\-ded from below.
With the free energy $W$ defined by
\be \label{free}
   e^{W(v)} = \int \prod_{x\in\Lambda} d^N\Phi_x \;
   e^{-S(\Phi,v)}
\ee
we obtain the LCE as the Taylor expansion
\be \label{taylor}
  W(v) = \left. \exp{\left( \sum_{x,y} v_{xy} 
   \frac{\partial}{\partial \widehat v_{xy}} \right) } W(\widehat v)
   \right\vert_{\widehat v =0}
\ee
and the obvious generalization to connected correlations 
and susceptibilities
\be \label{chi2n}
 \chi_{2n} = \sum_{x_2,x_3,\dots ,x_{2n}}
  < \Phi^1_0 \Phi^1_{x_2} \Phi^1_{x_3} \cdots \Phi^1_{x_{2n}} >^{conn} .
\ee
The rapidly increasing combinatorical complexity
can be managed by the approrpiate graphical device.
The application of the LCE in field theory has been poineered
by L\"uscher and Weisz \cite{LW1}.
L\"uscher and Weisz studied a
lattice $\Phi^{4}$-theory close to its continuum limit
in four dimensions.
They succeeded in a 14th order computation of susceptibility series
for arbitrary quartic couplings.
 
Normally, these expansions are applied to models on a lattice that is
unbounded in all direction, and to the study of 2nd
order transitions. There are
recent new fields of application and generalization of LCE.
A major topic are
field theories at finite temperature \cite{thomas}. 
Here one dimension of the lattice
is compactified to a torus of length given by
the inverse temperature $L_0$ in lattice units.
If $L_0$ is an even number, equivalence classes of graphs
can be introduced that do not depend on $L_0$ and hence are
the same as for $L_0=\infty$.
There are then mainly two adjustments necessary 
in order to generalize LCE to this case. First,
imbedding numbers change.
The second point is more laborious. 
Comparing field theory models at zero and non-zero temperature, 
the shift of transition
regions typically is rather small in bare parameters, whereas critical
behaviour such as critical exponents in many instances undergo a
considerable change. This different critical behaviour has to be resolved 
by the different high order behaviour of the susceptibility series.
It amounts to an even higher order computation than on lattices
unbounded in all directions.
Graphs contributing to the expansion must be able to wind around
the temperature torus sufficiently often in order to pick 
up sufficient finite temperature dependence. 

We briefly mention two other recent generalizations of LCE.
The first one is to apply LCE in a finite volume and to utilize
finite size scaling analysis in order to study both 1st and
2nd order transitions, as well as tricritical behaviour
\cite{hildegard}.
This is the topic of the talk given by H.~Meyer-Ortmanns, these 
proceedings.
The second generalization applies to models
that couple fields at lattice sites which are not nearest 
neighbours \cite{andreas2}.

\section{ALGEBRA OF GRAPHS}

In the range of orders considered, the approximated 
computational costs of 
increasing the LCE series
by two orders are a factor of about 30-40. 
Fast algorithms have to be developed to achieve this
computation.
Some of the problems that
one is faced with are briefly described in the following.

A first step is to restrict the classes of graphs to be
considered and write
correlation functions in terms of quantities that
allow for expansions into those restricted graph classes,
such as 1PI ones. 
Furthermore, 1PI graphs are composed of 1VI graphs
${\cal S}$ and renormalized moments ${\cal Q}$
(one external vertex only).

Incidence matrices provide a convenient way to represent 
graphs algebraically. 
For a graph $\Gamma$ with $V$ vertices, the incidence matrix $I_\Gamma$
is the symmetric $V\times V$-matrix obtained as follows.
\begin{itemize}
\item Enumerate the vertices in any order.
\item $I_\Gamma(i,i)$ = number of external lines attached
to the vertex $v_i$.
\item For $i\not= j$, $I_\Gamma(i,j)$ = number of common lines of the vertices
$v_i$ and $v_j$
\end{itemize}
As an example, the graph
%
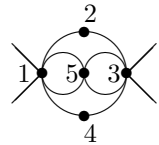
\begin{figure}[htb]
\setlength{\unitlength}{0.8cm}
\begin{picture}(4.0,0.8)
\put(3.8,0.6){\circle{1.4}}
\put(3.45,0.6){\circle{0.7}}
\put(4.15,0.6){\circle{0.7}}
\put(3.5,0.6){\makebox(0.2,0){$5$}}
\put(3.8,0.6){\circle*{0.16}}
\put(2.7,0.6){\makebox(0.2,0){$1$}}
\put(3.1,0.6){\circle*{0.16}}
\put(4.2,0.6){\makebox(0.2,0){$3$}}
\put(4.5,0.6){\circle*{0.16}}
\put(3.8,1.6){\makebox(0.2,0){$2$}}
\put(3.8,1.3){\circle*{0.16}}
\put(3.8,-0.4){\makebox(0.2,0){$4$}}
\put(3.8,-0.1){\circle*{0.16}}
\put(4.5,0.6){\line(1,1){0.5}}
\put(4.5,0.6){\line(1,-1){0.5}}
\put(3.1,0.6){\line(-1,1){0.5}}
\put(3.1,0.6){\line(-1,-1){0.5}}
\end{picture}
\end{figure}

\noindent
is represented by the incidence matrix
\[
   I_\Gamma \; = \; \left( 
   \begin{array}{ ccccc }
    2 & 1 & 0 & 1 & 2 \\
      & 0 & 1 & 0 & 0 \\
      &   & 2 & 1 & 2 \\
   \multicolumn{3}{c}{\mbox{symm.}} & 0 & 0 \\
      &   &   &   & 0 \\
   \end{array} \right)
\]
This way of representing graphs is not unique because
the vertices can be enumerated arbitrarily.
Comparison of two graphs needs
huge factorial numbers of simultaneous
permutations of rows and columns.
The actual construction of graphs is done 
by iteration on the number
of lines. 
Equivalent graphs have to be kept only once,
and the number of comparisions is a measure of the efficiency
of the algorithms.

A "canonical" representation of a graph $\Gamma$ that is
both unique and efficient is the following.
Introduce a (partial) order relation on the set of vertices
and enumerate the vertices in accordance with it.
The incidence matrix then is defined by
\be
    I^{can}_\Gamma (i,j) = \max_{\pi\in\Pi^\prime} I_\Gamma (\pi(i),\pi(j)),
\ee
where $\Pi^\prime$
is the subset of permutations that exchange vertices
only if they stay unordered relative to each other under
the preordering. The maximum is with respect to any total order
relation on symmetric matrices.

The preorder of vertices should be as complete as possible.
If all "local" properties of vertices are taken into account
in an optimal way we obtain the order relation
introduced by L\"uscher and Weisz. They succeeded in a 14th
order computation for 2- and 4-point functions.
Even higher orders need to take into account more global
properties. A good example is given by
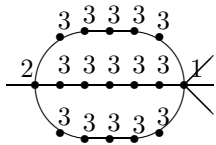
\begin{figure}[htb]

\begin{center}
\setlength{\unitlength}{0.55cm}
\begin{picture}(4.0,0.7)

\put(1.9000,0.8000){\oval(3.600,2.600)}
\put(-0.2000,1.2000){\makebox(0.2,0){$2$}}
\put(0.1000,0.8000){\circle*{0.16}}
\put(3.9000,1.2000){\makebox(0.2,0){$1$}}
\put(3.7000,0.8000){\circle*{0.16}}

\put(0.7000,2.3700){\makebox(0.2,0){$3$}}
\put(0.7000,1.9700){\circle*{0.16}}
\put(1.3000,2.5000){\makebox(0.2,0){$3$}}
\put(1.3000,2.1000){\circle*{0.16}}
\put(1.9000,2.5000){\makebox(0.2,0){$3$}}
\put(1.9000,2.1000){\circle*{0.16}}
\put(2.5000,2.5000){\makebox(0.2,0){$3$}}
\put(2.5000,2.1000){\circle*{0.16}}
\put(3.1000,2.3700){\makebox(0.2,0){$3$}}
\put(3.1000,1.9700){\circle*{0.16}}

\put(0.7000,0.0300){\makebox(0.2,0){$3$}}
\put(0.7000,-0.3700){\circle*{0.16}}
\put(1.3000,-0.1000){\makebox(0.2,0){$3$}}
\put(1.3000,-0.5000){\circle*{0.16}}
\put(1.9000,-0.1000){\makebox(0.2,0){$3$}}
\put(1.9000,-0.5000){\circle*{0.16}}
\put(2.5000,-0.1600){\makebox(0.2,0){$3$}}
\put(2.5000,-0.5000){\circle*{0.16}}
\put(3.1000,0.0300){\makebox(0.2,0){$3$}}
\put(3.1000,-0.3700){\circle*{0.16}}
\put(0.1000,0.8000){\line(1,0){3.6000}}
\put(0.7000,1.3000){\makebox(0.2,0){$3$}}
\put(0.7000,0.8000){\circle*{0.16}}
\put(1.3000,1.3000){\makebox(0.2,0){$3$}}
\put(1.3000,0.8000){\circle*{0.16}}
\put(1.9000,1.3000){\makebox(0.2,0){$3$}}
\put(1.9000,0.8000){\circle*{0.16}}
\put(2.5000,1.3000){\makebox(0.2,0){$3$}}
\put(2.5000,0.8000){\circle*{0.16}}
\put(3.1000,1.3000){\makebox(0.2,0){$3$}}
\put(3.1000,0.8000){\circle*{0.16}}
\put(0.1000,0.8000){\line(-1,0){0.7}}
\put(3.7000,0.8000){\line(1,0){0.7}}
\put(3.7000,0.8000){\line(1,1){0.7}}
\put(3.7000,0.8000){\line(1,-1){0.7}}

\end{picture}
\end{center}

\end{figure}

\noindent
with many 2-vertices staying unordered. Taking into account 
properties like the distance 
to vertices that have more than 2 lines attached, 
vertex ordering can be enhanced to give

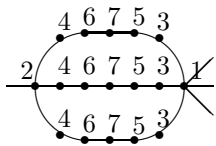
\begin{figure}[htb]

\begin{center}
\setlength{\unitlength}{0.55cm}
\begin{picture}(4.0,0.7)

\put(1.9000,0.8000){\oval(3.600,2.600)}
\put(-0.2000,1.2000){\makebox(0.2,0){$2$}}
\put(0.1000,0.8000){\circle*{0.16}}
\put(3.9000,1.2000){\makebox(0.2,0){$1$}}
\put(3.7000,0.8000){\circle*{0.16}}

\put(0.7000,2.3700){\makebox(0.2,0){$4$}}
\put(0.7000,1.9700){\circle*{0.16}}
\put(1.3000,2.5000){\makebox(0.2,0){$6$}}
\put(1.3000,2.1000){\circle*{0.16}}
\put(1.9000,2.5000){\makebox(0.2,0){$7$}}
\put(1.9000,2.1000){\circle*{0.16}}
\put(2.5000,2.5000){\makebox(0.2,0){$5$}}
\put(2.5000,2.1000){\circle*{0.16}}
\put(3.1000,2.3700){\makebox(0.2,0){$3$}}
\put(3.1000,1.9700){\circle*{0.16}}

\put(0.7000,0.0300){\makebox(0.2,0){$4$}}
\put(0.7000,-0.3700){\circle*{0.16}}
\put(1.3000,-0.1000){\makebox(0.2,0){$6$}}
\put(1.3000,-0.5000){\circle*{0.16}}
\put(1.9000,-0.1000){\makebox(0.2,0){$7$}}
\put(1.9000,-0.5000){\circle*{0.16}}
\put(2.5000,-0.1600){\makebox(0.2,0){$5$}}
\put(2.5000,-0.5000){\circle*{0.16}}
\put(3.1000,0.0300){\makebox(0.2,0){$3$}}
\put(3.1000,-0.3700){\circle*{0.16}}
\put(0.1000,0.8000){\line(1,0){3.6000}}
\put(0.7000,1.3000){\makebox(0.2,0){$4$}}
\put(0.7000,0.8000){\circle*{0.16}}
\put(1.3000,1.3000){\makebox(0.2,0){$6$}}
\put(1.3000,0.8000){\circle*{0.16}}
\put(1.9000,1.3000){\makebox(0.2,0){$7$}}
\put(1.9000,0.8000){\circle*{0.16}}
\put(2.5000,1.3000){\makebox(0.2,0){$5$}}
\put(2.5000,0.8000){\circle*{0.16}}
\put(3.1000,1.3000){\makebox(0.2,0){$3$}}
\put(3.1000,0.8000){\circle*{0.16}}
\put(0.1000,0.8000){\line(-1,0){0.7}}
\put(3.7000,0.8000){\line(1,0){0.7}}
\put(3.7000,0.8000){\line(1,1){0.7}}
\put(3.7000,0.8000){\line(1,-1){0.7}}

\end{picture}
\end{center}

\end{figure}

\noindent
There are $(3!)^5$ permutations left to build $I^{can}_\Gamma$
instead of $(15!)$.
Furthermore, order relations can be iteratively improved
by taking into account order numbers of neighboured
vertices, next to nearest neighbours and so one.
Normally, one or two iterations are sufficient.
 
The table shows the number of graphs for some classes
generated as yet.
The index $k=2,4,6$ denotes the number of external lines
attached, $L$ is the order (number of internal lines).
 
The contribution of a graph, its weight, 
consists of two factors. The first 
factor is a
rational number, itself a product of internal and (inverse) 
topological symmetry numbers and of the lattice imbedding
number. It is only the latter that depends on the lattice size.
The second factor comes from the product over the vertices. 
Every vertex contributes a finite
dimensional integral that depends on the number of lines
entering the vertex. The complete weight is a real number.
The accumulation of roundoff errors is avoided by introducing 
so-called vertex structures.
A vertex structure of a graph $\Gamma$ is the multitupel of non-negative
integers,
$\Gamma \to (\mu_1,\mu_2,\ldots )$,
where $\mu_i$ denotes the number of vertices
of $\Gamma$ that have
precisely $i$ lines attached. 
Before multiplying the real coupling constants of the vertices,
graphs with identical vertex structure
are summed according to their rational weights.
 
An application to finite temperature field theory 
can be found in \cite{thomas2}.

\vskip 0.2cm
  
\noindent
\begin{tabular}{rrrrr}
\hline \hline
$L$ & $\cQ_2(L)$ & 
$\cS_2(L)$ & $\cS_4(L)$ & $\cS_6(L)$ \\
[0.5ex] \hline 

 0 & 1 & 1 & 1 & 1 \\
 1 & 0 & 0 & 0 & 0 \\
 2 & 1 & 0 & 1 & 1 \\
 3 & 0 & 1 & 1 & 2 \\
 4 & 4 & 0 & 4 & 6 \\
 5 & 0 & 2 & 4 & 11 \\
 6 & 15 & 3 & 20 & 46 \\
 7 &  0 & 8 & 27 & 91 \\
 8 & 79 & 9 & 117 & 349 \\
 9 &  0 & 40 & 214 & 837 \\
10 & 439 & 68 & 815 & 3140 \\ 
11 &   0 & 247 & 1830 & 8401 \\
12 & 2877 & 470 & 6721 & 31187 \\
13 &    0 & 1779 & 17028 & 90599 \\
14 & 20507 & 3937 & 61653 & 336582 \\ 
15 &    0 & 14801 & 170923 & 1042392 \\
16 & 161459 & 35509 & 621191 & 3895341 \\
17 &     0 & 135988 & 1834324 &  \\
18 & 1376794 & 350614 & 5548427 &  \\
19 &     0 & 1361878 & 20967387 &  \\ 
20 & 12693105 & 3705467 &  & \\ \hline
\end{tabular}

\end{document}